\documentstyle[floats,aps,aipbook,psfig]{revtex}

\newbox\grsign \setbox\grsign=\hbox{$>$} 
\newdimen\grdimen \grdimen=\ht\grsign
\newbox\laxbox \newbox\gaxbox
\setbox\gaxbox=\hbox{\raise.5ex\hbox{$>$}\llap
     {\lower.5ex\hbox{$\sim$}}}\ht1=\grdimen\dp1=0pt
\setbox\laxbox=\hbox{\raise.5ex\hbox{$<$}\llap
     {\lower.5ex\hbox{$\sim$}}}\ht2=\grdimen\dp2=0pt
\def\gax{\mathrel{\copy\gaxbox}}

\begin{document}

\title{A Comprehensive Search for Low-Energy Lines in BATSE GRBs}

\author{Michael S. Briggs$^1$, David L. Band$^2$, Robert D. Preece$^1$,  \\
Geoffrey N. Pendleton$^1$, William S. Paciesas$^1$,  \\
Lyle Ford$^2$, James L. Matteson$^2$}

\address{$^1$Department of Physics \\
         University of Alabama in Huntsville, Huntsville, AL 35899 \\
       $^2$Center for Astrophysics and Space Science \\
       University of California, San Diego, La Jolla, CA 92037-0111}

\maketitle

\begin{abstract}
A computer-based technique has been developed to search 
bright BATSE gamma-ray bursts for spectral lines in a comprehensive manner.
The first results of the search are discussed and an example line candidate
shown.
\end{abstract}

\section*{Motivation}

Prior to the launch of CGRO, and during the early days of the mission,
the BATSE team expected that GRB lines would be common and easily identified.
Our expectations were based upon the observations of KONUS 
\cite{Maz81},              
HEAO A-4 {\cite{Hue87} and {\it Ginga} \cite{Fen88,Mur88}.
Our early efforts were based upon these expectations: we searched
bright bursts visually,
examining consecutive spectra and also
the spectrum of the entire burst \cite{Pal94}.
Our expectations were not met: no significant lines have been found
\cite{Pal94,Ban96A}.

While the failure to find lines in BATSE GRBs does not yet imply a serious
discrepancy between the {\it Ginga} and BATSE results \cite{Pal94,Ban94}, it
is disappointing.
Previously, lines were interpreted as strongly supporting the theory that
GRBs originate nearby from a disk population.     
Results from BATSE on the spatial distribution of GRBs demonstrate that only
a minority of the bursts can originate from a disk population
\cite{Mee92,Hak94,Bri96A},
so if the existence of lines is confirmed, their properties might indicate the
presence of a disk subclass or give clues to the physics of
halo or cosmological sources.

Is the failure to detect lines in the BATSE data because lines do not 
exist or is the failure due to some inadequacy in the visual search approach?
To answer these questions a more systematic search is needed.
We have therefore implemented an automatic, computer-based comprehensive
line search.
The goal of the search is comprehensiveness---a brute force approach is
taken so that no significant line will be missed.
The purpose of the search is to identify line candidates---the search
need not perfectly evaluate their significances, which will be done 
later under direct human control.
The computer-based search has several advantages 
in addition to comprehensiveness:
subjectivity is eliminated,
there is no variation in detection
threshold due to human exhaustion,
there is no bias towards absorption or emission features,
and the search will collect statistics on all trial lines.

\section*{Comprehensive Search Method}

The addition to the BATSE instrument of the eight Spectroscopy Detectors (SDs)
was largely motivated by the KONUS results.
Simulations \cite{Ban95,Fre93}
and tests \cite{Pac96} of the performance of the BATSE 
SDs show that they should be capable of detecting KONUS or
{\it Ginga}-like GRB lines.
Each SD is a NaI(Tl) scintillator crystal, 12.7 cm diameter and 7.6 cm thick, 
viewed by a photomultiplier tube of the same diameter.  
To limit the effort to a manageable level, the initial searches are restricted
to lines below 100 keV,
so only detectors in high-gain mode
are useful.  Typically, such detectors
cover the energy range of about 10 keV to about 1400 keV, although
analysis begins at about 15 or 20 keV due to an electronic artifact
\cite{Ban92}.  Work to model this artifact continues, so that 
in the future we will be able to extend the analysis to lower energies.
Analysis of the data is based upon an instrument 
response model which accounts for the
detectors, nearby spacecraft material, and scattering from the Earth's
atmosphere \cite{Pen95}. 
When a burst occurs, high-time resolution SHERB (Spectroscopy High
Energy Resolution Burst) data are collected using a time-to-spill algorithm.

Since we do not know a priori when or how long a line will exist, we
search essentially all time scales available.
We search each single SHERB record, every consecutive pair, triple, and 
group of 4, 5, 6, 8, 10, 14, 20, \ldots records and the sum of all the SHERB
records. 
Clearly, many overlapping intervals are searched, and thus the
searched intervals are not all independent.
Additionally, many intervals will have insufficient 
signal-to-noise ratio for a real line
to be detectable---these intervals serve as controls.
While not every possible consecutive combination of records is formed,
enough combinations are searched so that no line should be missed, although
its significance might not be optimized.
           
Similarly, we do not know a priori at what energies lines occur.
Therefore, we first fit each spectrum with a continuum model, and then we
perform continuum plus line fits using a closely spaced, fixed grid of trial 
centroids.    
Since the trial centroids are separated by one-third
the detector resolution FWHM, no line candidate should be missed, although
the exact centroid will not be found by the search.

Candidates are identified by large ($\gax 20$) changes in $\chi^2$.
We have switched to using $\Delta \chi^2$ instead of the F-test because
it is more appropriate when the errors are known
\cite{Lam95,Ead71,Ban96B}.
After a candidate is identified, other time intervals will be tried and
the centroid will be made a free parameter in the fits.

To increase the robustness of the automatic nonlinear fits, we use a continuum
model with a small number of parameters, namely the Comptonized model, which is
 a power-law times an exponential cutoff.    This is fit to 
a restricted energy range of $\approx 400$ keV.
Our initial search is confined to lines which are narrow compared
to the detector resolution and
we use the simplest line model: an additive (or subtractive) Gaussian.
So that there is continuum at energies below any line, the lowest energy
trial centroid is two detector-resolution FWHM above the starting energy
of the fits.   Since very few of the fits begin below 15 keV, only rarely
do we search for 20 keV lines.

We show as an example of a burst without a significant line the data of
SD~2 for GRB 920627.
The 63 SHERB records of SD~2 are summed into 693 spectra which are
fit with trial lines at 20 different centroids from 27.3 to 100 keV.
The largest change in $\chi^2$ obtained by adding a line was 13.3
(Fig.~1),
which has a chance probability in a single fit of a few tenths of a percent.
Considering the large number  of spectra
searched (albeit not all independent), this is insignificant.

\begin{figure}[tb!]              
\mbox{
\psfig{figure=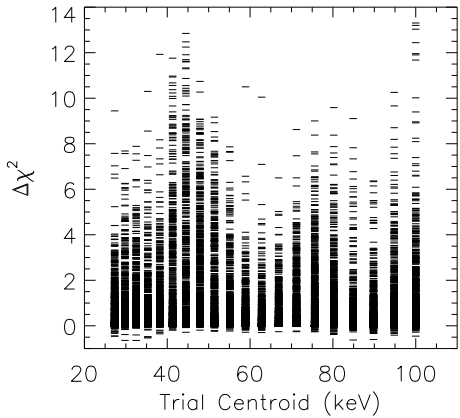,width=61mm,%
bbllx=106bp,bblly=100bp,bburx=525bp,bbury=490bp,clip=} 
\hspace*{2.5mm}
\begin{minipage}[b]{50mm}
\caption{Results of the comprehensive line search for GRB 920627 (trigger
1676), SD~2.    The figure shows
the change in $\chi^2$ resulting from adding to the continuum model a 
narrow, additive Gaussian line.
While SD~2 has useable data from 15--1240 keV, the search was made 
using the Comptonized continuum model and the restricted energy 
range 15--390 keV.
The few points with $\Delta \chi^2 < 0$ are due to poor convergence.
\protect\vspace{7.0mm}
}
\end{minipage}
}
\end{figure}

\section*{First Results of the Comprehensive Search}

The search results for GRB 940703 are quite different from those for 
GRB~920627.
Only one high gain detector, SD~5, viewed the burst at an angle less than
$100^\circ$.
The data from SD~5 comprises
79 SHERB records, from which 917 spectra were formed.
The largest value of $\Delta \chi^2$ identified in the search was 58.2
and many overlapping intervals had $\Delta \chi^2$ values approaching this
value (see Fig.~2).
When more careful fits are made, using the Band spectral form
\cite{Ban93} and
the entire available energy range, the candidate is found to have
$\Delta \chi^2$ of ``only'' 23.4.
The difference in the values of $\Delta \chi^2$ is because for this very bright
burst the Comptonized model is inadequate even over a restricted energy range.

Most of the line significance appears to come from a shorter interval
which excludes periods of weak emission, for which $\Delta \chi^2$ = 28.1.
The spectrum of this interval is shown in Fig.~3---the candidate is
seen to be an {\it emission} feature with an equivalent width of 2.1 keV.
The line is narrow---allowing the width to vary does not significantly
reduce $\chi^2$.
For 3 additional parameters,
the chance probability in any particular fit of such
a large $\Delta \chi^2$, if no line actually exists, is $3\times10^{-6}$.
Additionally, the residual plots show that adding the line to the model
eliminates conspicuous runs of high, then low, residuals.
Even considering the large number of spectra searched, the line is
highly significant---the number of effectively independent spectra is
well below 917 and the line is significant in more than one spectrum.


\begin{figure}[tb!]              
\mbox{
\psfig{figure=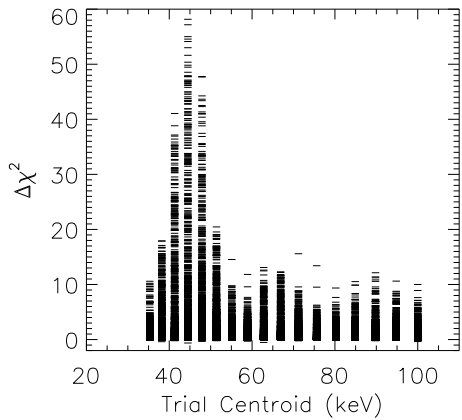,width=61mm,%
bbllx=106bp,bblly=100bp,bburx=525bp,bbury=490bp,clip=} 
\hspace*{3mm}
\begin{minipage}[b]{50mm}
\caption{Results of the comprehensive line search for GRB~940703 (trigger
3057), SD~5.    The figure shows
the change in $\chi^2$ resulting from adding to the continuum model a 
narrow, additive Gaussian line.
While SD~5 has useable data from 20--1380 keV, the search was made
using the Comptonized continuum 
model and the restricted energy range 20--397 keV.
The angle of the detector
normal with the burst was 51$^\circ$ and with the geocenter, 79$^\circ$.
The few points with $\Delta \chi^2 < 0$ are due to poor convergence.
\protect\vspace{1.5mm}
}
\end{minipage}
}
\end{figure}

\begin{figure}[tb!]          
\mbox{             
\psfig{figure=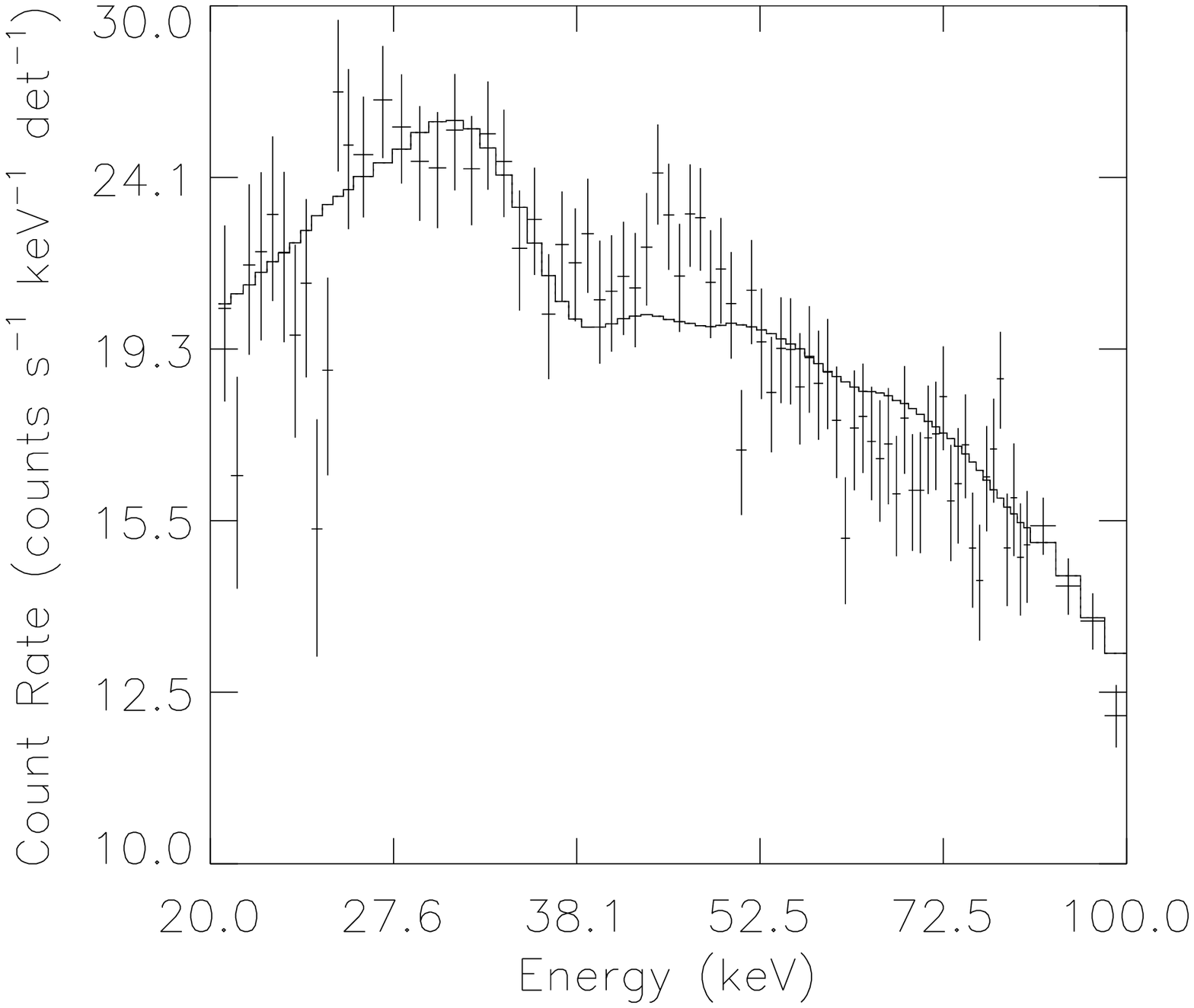,width=59.4mm,%
bbllx=15bp,bblly=225bp,bburx=560bp,bbury=645bp,clip=}
\hspace{0.25mm}
\psfig{figure=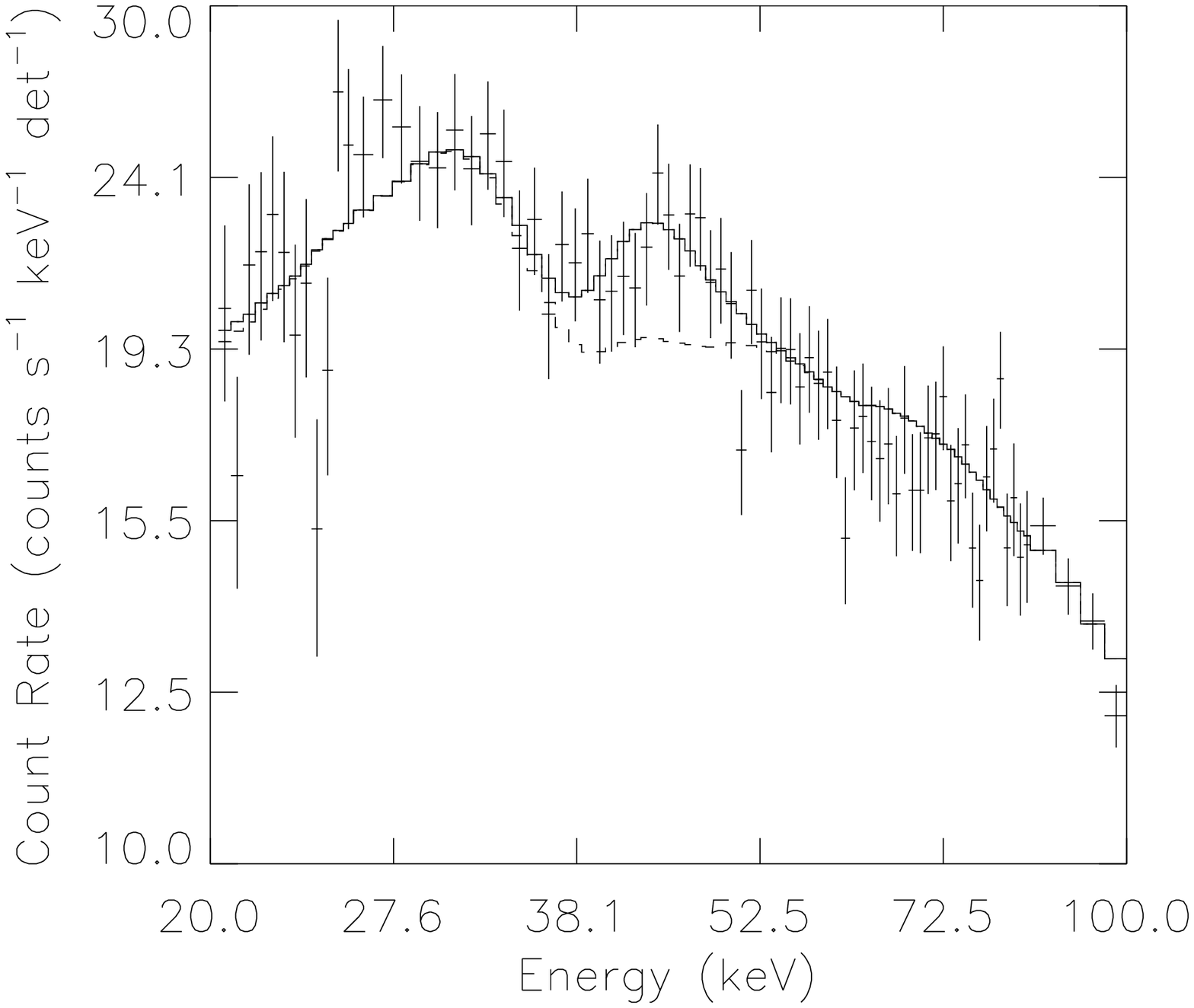,width=54.0mm,%
bbllx=64bp,bblly=225bp,bburx=560bp,bbury=645bp,clip=} }
\vspace{0.1mm}

\mbox{
\psfig{figure=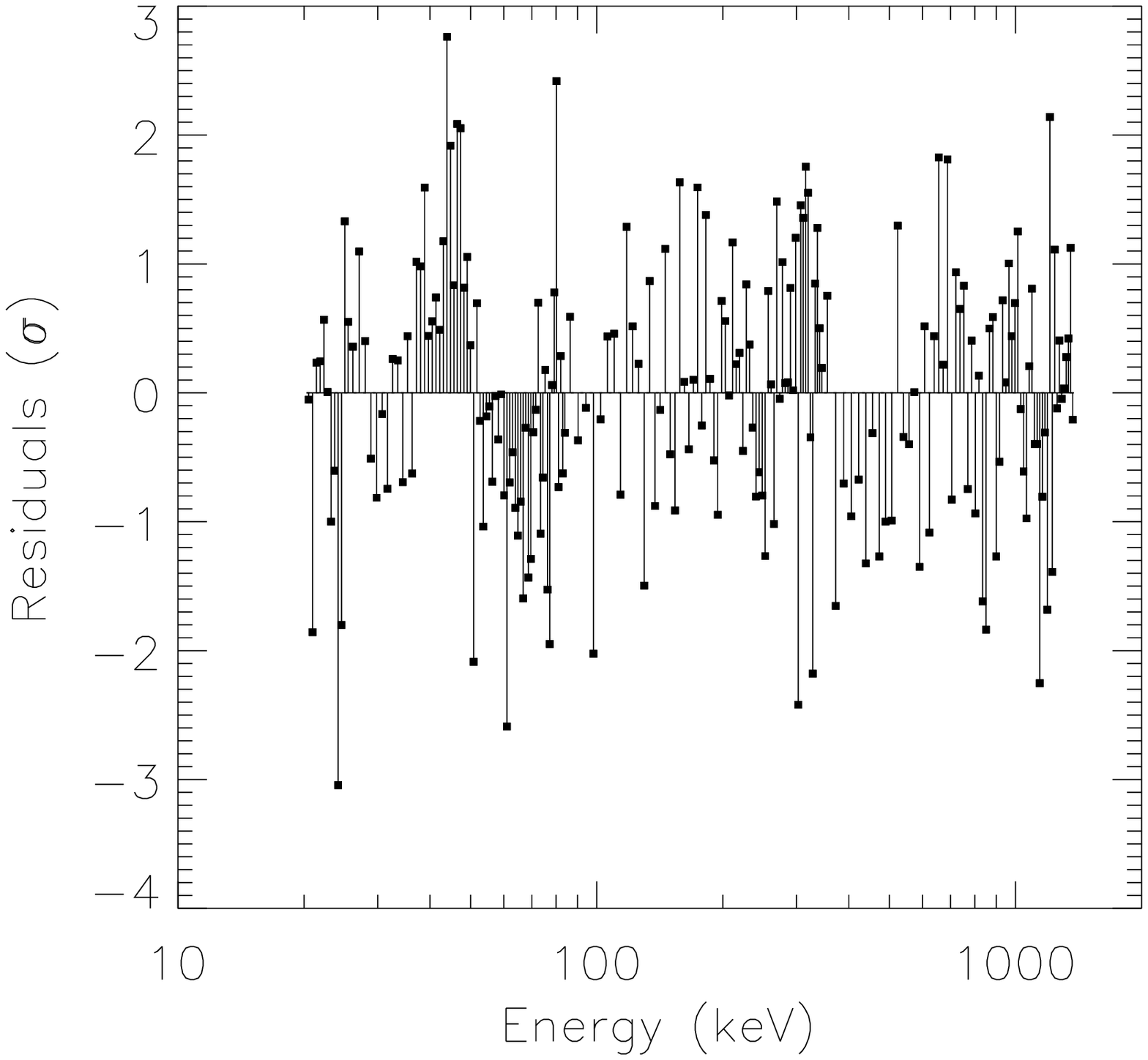,width=59.4mm,%
bbllx=15bp,bblly=190bp,bburx=560bp,bbury=645bp,clip=}
\hspace*{2.45mm}
\psfig{figure=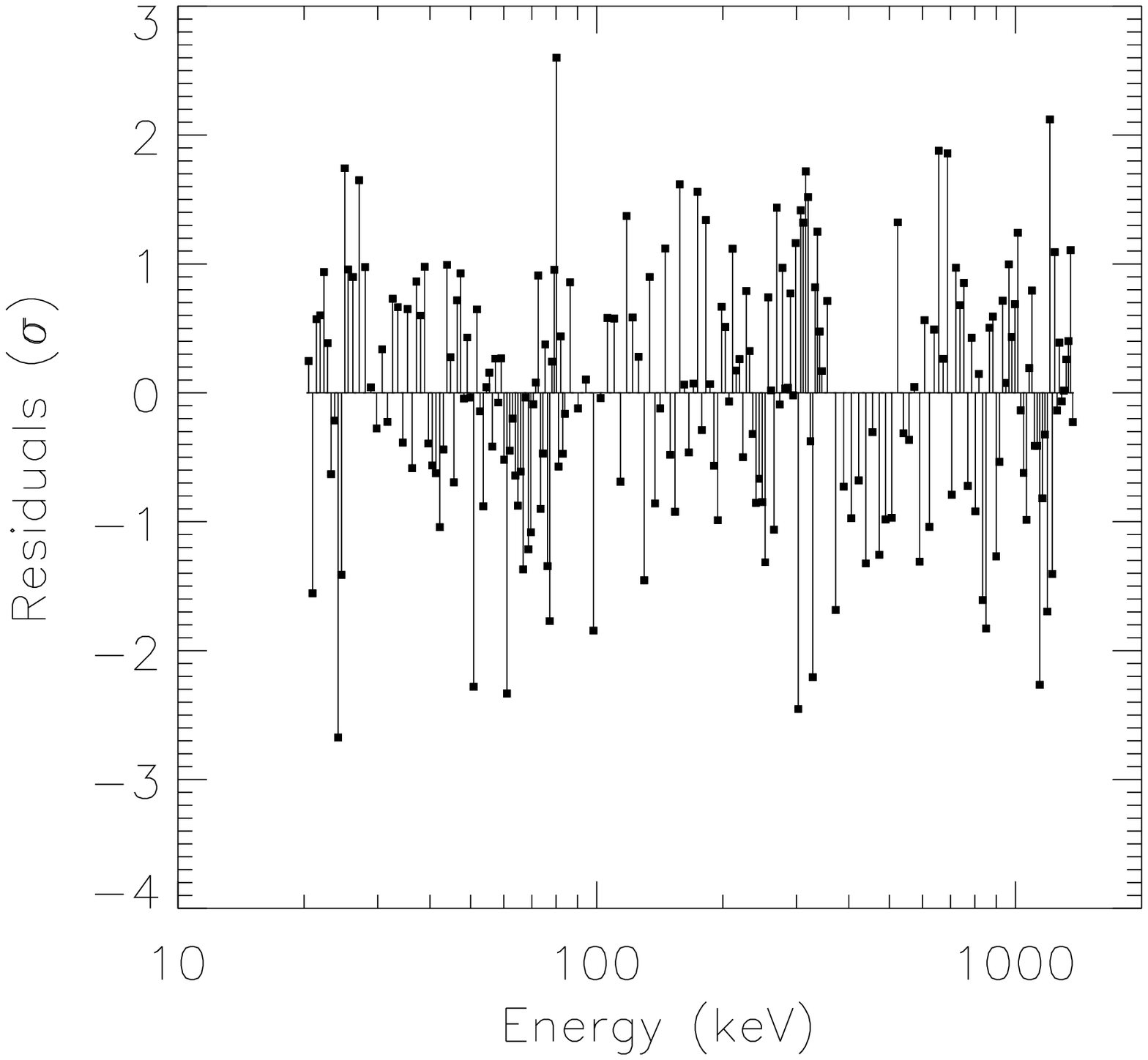,width=51.9mm,%
bbllx=84bp,bblly=190bp,bburx=560bp,bbury=645bp,clip=} }

\caption{Data from SD~5, GRB 940703, for the 
interval 42.112--64.704~s.
Top panels: Count rate data and fits, shown up to 100 keV.
Bottom panels: Count rate residuals (data $-$ model) in units of $\sigma$
for the entire energy range used in the fits.
Left panels: A continuum-only fit, using the Band spectral form;
$\chi^2$ = 212.0 with 199 degrees-of-freedom.
Right panels: Added to the fits is an narrow, additive Gaussian line;
$\chi^2$ = 183.9 with 197 degrees-of-freedom.
The ``hump'' at 30 keV in the data and continuum model is expected
from the detector physics \protect\cite{Bri96B,Pac96}.
The contribution of atmospheric scattering to the counts in the line region
is $\approx$4\%.}
\end{figure}

To date, 42 of the brightest GRBs have been searched by this technique.
There was an average of 2.1 detectors per burst and 53 SHERB records per
detector, for a total of 4729 SHERB records.   From these records
51,651 spectra were formed and 861,372 fits performed.     
Partial examination of the results has identified 8 candidates with
$\Delta \chi^2 > 20$.    There are both absorption and emission 
candidates and their centroids range from 40 to 70 keV.


More work remains to be done.
In most cases several high-gain SDs observe a gamma-ray burst.
We can therefore usually test the reality of a line candidate by examining
the data of the other detectors.
If the lines are real, in all cases the data should be consistent and
in some cases there should be confirmation: statistically significant
detections in more than one detector.
The ability to perform these tests is a major advantage of BATSE, but
it is also time-consuming because there are many issues.
Until we have completed these tests,
the BATSE team considers the features identified by the
comprehensive search to be line candidates
rather than detections.

\end{document}